\documentclass[
superscriptaddress,
twocolumn,
amsmath,amssymb,
aps,
prl,
longbibliography
]{revtex4-2}

\usepackage{graphicx}
\usepackage{dcolumn}
\usepackage{xcolor}
\usepackage{bm}
\usepackage{physics}
\usepackage{hyperref}
\usepackage{comment}
\hypersetup{
	colorlinks = true,
	urlcolor   = blue,
	linkcolor  = blue,
	citecolor  = red
}
\usepackage{algorithm}
\usepackage{algpseudocode}
\usepackage{float}

\makeatletter
\let\newfloat\newfloat@ltx
\makeatother

\begin{document}

\title{Bayesian stepwise estimation of qubit rotations}
\author{M. Manrique}
\affiliation{Dipartimento di Scienze, Universit\`a degli Studi Roma Tre, Via della Vasca Navale, 84, 00146 Rome, Italy}
\author{M. Barbieri}
\affiliation{Dipartimento di Scienze, Universit\`a degli Studi Roma Tre, Via della Vasca Navale, 84, 00146 Rome, Italy}
\affiliation{Istituto Nazionale di Ottica - CNR, Largo E. Fermi 6, 50125 Florence, Italy}
\affiliation{INFN, Sezione di Roma Tre, Via Della Vasca Navale 84, 00146 Roma, Italy}
\author{A. Di Vizio}
\affiliation{Dipartimento di Scienze, Universit\`a degli Studi Roma Tre, Via della Vasca Navale, 84, 00146 Rome, Italy}
\author{M. Parisi}
\affiliation{Dipartimento di Scienze, Universit\`a degli Studi Roma Tre, Via della Vasca Navale, 84, 00146 Rome, Italy}
\author{G. Bizzarri}
\affiliation{Dipartimento di Scienze, Universit\`a degli Studi Roma Tre, Via della Vasca Navale, 84, 00146 Rome, Italy}
\author{I. Gianani}
\affiliation{Dipartimento di Scienze, Universit\`a degli Studi Roma Tre, Via della Vasca Navale, 84, 00146 Rome, Italy}
\author{M.G.A. Paris}
\affiliation{Dipartimento di Fisica ``Aldo Pontremoli", Universit\`a degli Studi di Milano, Via Celoria 16, 20133 Milan, Italy}

\date{\today}

\begin{abstract}
This work investigates Bayesian stepwise estimation (Se) for measuring the
two parameters of a unitary qubit rotation. While asymptotic analysis
predicts a precision advantage for SE over joint estimation (JE) in regimes
where the quantum Fisher information matrix is near-singular (``sloppy"
models), we demonstrate that this advantage is mitigated within a practical
Bayesian framework with limited resources. We experimentally implement a SE
protocol using polarisation qubits, achieving uncertainties close to the
classical Van Trees bounds. However, comparing the total error to the
ultimate quantum Van Trees bound for JE reveals that averaging over prior
distributions erases the asymptotic SE advantage. Nevertheless, the
stepwise strategy retains a significant practical benefit as it operates
effectively with simple, fixed measurements, whereas saturating the JE
bound typically requires complex, parameter-dependent operations.
\end{abstract}
\maketitle

\section{Introduction}
Quantum metrology exploits the unique properties of quantum systems to estimate physical parameters with precision beyond classical limits~\cite{Giovannetti2006,Liu2020,Degen2017,Paris2009}. While the ultimate precision bounds for single-parameter
estimation are well-established, their multiparameter counterparts present
relevant and interesting challenges~\cite{Szczykulska2016,Demkowicz2020}. The natural approach of joint
estimation (JE) is fundamentally limited by two phenomena. The first is
quantum in nature, measurement incompatibility, since the optimal
measurements for different parameters may not commute, preventing the
simultaneous saturation of the quantum Cramer-Rao bound (QCRB)~\cite{Ragy2016,Albarelli2020}. 

The second, sloppiness, is a form of classical indeterminacy (or lack of parameter identifiability) that occurs when parameters are not independently encoded, leading to a singular or near-singular Quantum Fisher Information Matrix (QFIM)~\cite{Waterfall2006,Goldberg2021}. In  sloppy
models, the statistical model depends mostly or only on a combination of
the parameters, making their individual estimation nearly or fully impossible.

To circumvent these limitations, stepwise estimation (SE) has been recently
proposed
as a powerful alternative framework~\cite{Mukhopadhyay2025,fazio2025orders}. In SE, the total resources are
divided to estimate parameters sequentially: a subset of measurements is
used to obtain an estimate for the first parameter, which is then used as
prior knowledge to inform the estimation of the second parameter in the
remaining rounds. Analytically, it has been shown that this strategy can
achieve a total error that is strictly lower than the JE
bound  in regimes where the QFIM is nearly singular~\cite{Mukhopadhyay2025}. This advantage is
not merely theoretical, Bayesian implementations across diverse platforms,
including qubit probes, three-level Landau-Zener systems, and critical
many-body Ising chains, where SE can restore a quantum-enhanced scaling
advantage often lost in JE due to singularities~\cite{Mukhopadhyay2025}. These findings motivate the experimental exploration of SE protocols in realistic scenarios, where resources and prior knowledge are finite.

Concurrently, the fundamental trade-off between sloppiness and
incompatibility has been analyzed in two-parameter qubit models
incorporating a tunable "scrambling" operation~\cite{He2025}. Results imply that
minimizing sloppiness to enhance precision necessarily maximizes
incompatibility, and vice versa. Notably, it has been found that for such a qubit
model, the asymptotically achievable Holevo bound and the Nagaoka bound
(saturable with separable measurements) coincide. This result
underscores that the ultimate precision limits for both joint and stepwise
strategies are fundamentally set by the degree of sloppiness in the model~\cite{He2025}.

In this work, we bridge the gap between these analytical asymptotic bounds
and practical experimental implementation. We present a Bayesian stepwise
estimation protocol for estimating the two parameters of a unitary rotation
on a photonic qubit. While the asymptotic analysis predicts an advantage
for SE in sloppy regions, the transition to the Bayesian framework with
limited resources and finite prior knowledge introduces new considerations.
In our protocol, one parameter is first estimated, and its posterior distribution is then used as prior information for the estimation of the second parameter.

We experimentally implement this protocol using polarisation qubits and
find that the achieved uncertainties are close to the classical Van Trees
bounds for the stepwise strategy. However, when comparing the total error
to the ultimate quantum Van Trees bound for JE, we observe
that the potential advantage of SE is mitigated within the Bayesian
framework. We conclude that while the asymptotic, local CRB analysis
suggests clear regimes where SE outperforms JE, the practical Bayesian
implementation averages over prior distributions, often erasing this
advantage. Nevertheless, the stepwise strategy retains a significant
practical benefit: it operates effectively with simple, fixed projective
measurements (e.g., a $Z$-basis measurement), whereas saturating the JE bound would typically require complex, parameter-dependent, and
often infeasible collective measurements.

Our results provide a crucial reality check on the application of
multiparameter quantum metrology theorems, highlighting the interplay
between fundamental quantum bounds, practical measurement constraints, and
the role of prior information in Bayesian inference.

\section{Results}

The system we investigate consists in a collection of qubits undergoing a transformation characterised by a pair of parameters, the estimation of which is our final goal. This is in line with previous efforts aiming at the optimising resources in the assessment of quantum processes~\cite{Rozema2014,Zhou2015}. These qubits are encoded in the polarisation of heralded single photons. These are produced by means of parametric downconversion from a 3-mm nonlinear crystal (beta barium borate, BBO) illuminated by a pulsed pump ($\lambda_p=405$ nm, rep. rate 80 MHz, average power 50 mW, obtained from second harmonic generation from a 1 mm BBO). Photons are produced by energy-degenerate type I phase matching, spectrally filtered by an interference filter (full width half maximum 7.5 nm) and by a single-mode fibre.

Photons are prepared at the output of the fibre in the horizontal polarisation, and then reach a birefringent object, a quarter-wave plate for $\lambda_p$. This is set with its axis at an angle $\theta/2$ with respect to the horizonal, corresponding to the logic state $\ket{0}$, and imparts a phase shift $2\gamma$ between the slow and fast polarisations. Following this transformation, photons are measured in the logical basis, physically implemented by horizontal-vertical discrimination. 

The expression of the polarisation transformation is written as
\begin{equation}
    U(\theta,\gamma)=\exp[-i\gamma(\cos{\theta}\,\sigma_x+\sin{\theta}\,\sigma_z) ],
\end{equation}
where $\sigma_x$ and $\sigma_z$ are the first and third Pauli operators, defined with respect to the measurement basis. The two parameters to be estimated, based on the measurement statistics, are thus $\theta$ and $\gamma$. Knowing the expressions of the detection probabilities  $p_0=\vert \bra{0}U(\theta,\gamma)\ket{0}\vert^2$ and $p_1=\vert \bra{1}U(\theta,\gamma)\ket{0}\vert^2$, it is thus possible to infer the value of the parameters from the results. In formal terms, there exist functions mapping the collection of the experimental outcomes, here denoted $\vec x$, into estimated values $\hat \lambda_1$ and $\hat \lambda_2$. These functions are called estimators.

\begin{figure*}[ht!]
    \centering
    \includegraphics[width=\linewidth]{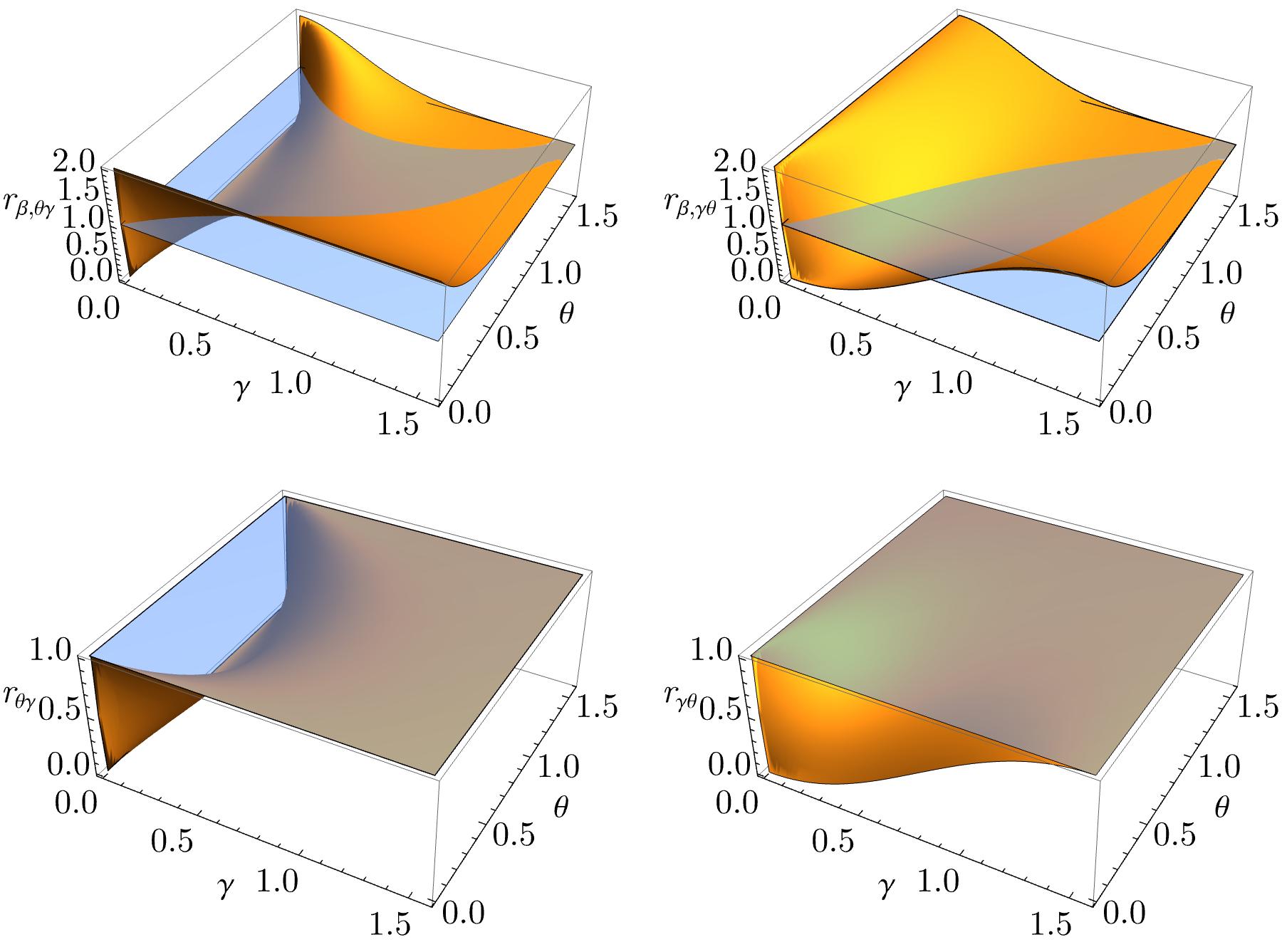}
    \caption{Performance of the SE, compared to JE, as captured by the quantity $r_\beta$, as well as its optimised version $r$. The theoretical predictions are reported as a function of the axis $\theta$ and of the phase shift $\gamma$ as the gold surface. The blue surface corresponds to the limit 1.  When resources are equally allocated ($\beta =1/2$), an advantage, identified by $r_\beta<1$, can be highlighted, depending on the estimation sequence. This is even more relevant when optimising over the allocation of resources.}
    \label{fig:ratio}
\end{figure*}

In the standard treatment of quantum metrology, when the two parameters are measured jointly, the minimal attainable variances on $\hat \lambda_1$ $\hat \lambda_2$ are bound through the QFIM $Q$ by means of the CRB:
\begin{equation}
\label{eq:CRB}
    \Sigma \geq \frac{1}{N}Q^{-1},
\end{equation}
where $\Sigma$ is the covariance matrix of the two parameters, and $N$ is the number of experimental runs. The bound \eqref{eq:CRB} implies scalar bounds on the individual uncertainty on $\lambda_i$ ($i=1,2$)
\begin{equation}
\label{eq:CRBind}
    \sigma_i^2 \geq \frac{1}{N}(Q^{-1})_{ii}. 
\end{equation}
In general, there is no guarantee that the variance at the parameter CRB can actually be achieved in a joint measurement. A tighter inequality is given by the Holevo-CRB, defined as
\begin{equation}
    \label{eq:HCRB}
    N \Tr({\Sigma})\geq N C_H=N\min_{\boldsymbol{X} } \Tr{\text{Re}[\mathcal{Z}(\boldsymbol{X})]+\|\text{Im}[\mathcal{Z}(\boldsymbol{X})]\|_1} .
\end{equation}
where $\boldsymbol{X}$ is a list of operators satisfying $\Tr[\partial_\mu\rho X_\nu]=\delta_{\mu\nu}$, $\mathcal{Z}_{\mu\nu}(\boldsymbol{X})=\Tr[\rho X_\mu X_\nu]$, and $\|A\|_1=\sqrt{A^\dag A}$. 
This is valid for an arbitrary measurement, even collective ones on multiple copies, however, in our case, separate ones are sufficient. In the SE approach, instead, the parameters are estimated one at the time in separate rounds. For the first parameter, the joint CRB is still relevant:  now we are already equipped with an estimate for $\lambda_1$, we can make use of single-parameter estimation for the second, thus the individual single-parameter CRBs hold:
\begin{equation}
\label{eq:sCRB}
    \sigma_2^2 \geq \frac{1}{(1-\beta) N Q_{ii}}, 
\end{equation}
where $\beta$ indicates the fraction of repetitions devoted to estimating $\lambda_1$. The performance of the stepwise approach can thus achieve the minimal variance
\begin{equation}
    \label{eq:sigmaCRB}
    N \Tr({\Sigma})\geq \frac{1}{\beta }(Q^{-1})_{11}+ \frac{1}{(1-\beta) Q_{22}},
\end{equation}
that is compared to the Holevo bound in order to appraise the relative merits of this strategy. For this purpose, we define the ratio
\begin{equation}
r_\beta = \frac{1}{C_H} \left(\frac{1}{\beta }(Q^{-1})_{11}+ \frac{1}{(1-\beta) Q_{22}}\right),
\end{equation}
as well as its optimised version $r =\min_\beta \,r_\beta$. These are plotted in Fig.~\ref{fig:ratio} for our case, with the two possible orderings of the estimations~\cite{fazio2025orders}: $r_{\theta\gamma}$ corresponds to $\theta$ first, then $\gamma$, and vice versa for $r_{\gamma\theta}$.
A better performance of the stepwise approach is expected in proximity of $\gamma\simeq 0$, {\it i.e.} when the transformation has little effect on the state. This regime is understood by looking at the determinant of the QFIM
\begin{equation}
    \det Q = 16 \sin^2\theta\sin^4\gamma.
\end{equation}
The available information in a JE tends to vanish in this limit.

The discussion above is fully relevant in the asymptotic regime of large $N$. In order to account for limited resources, the Bayesian perspective is particularly suitable, since it explicitly includes the {\it a priori} knowledge $A(\theta,\gamma)$ on the parameters, expressed by means of distributions. In our example we consider, for the sake of practicality, normal distributions $P_a(\lambda|\lambda_0,\tau)$ centred on $\lambda_0$ with the same variance $\tau^2$ for both parameters: $ A(\theta,\gamma)=P_a(\theta|\theta_0,\tau)P_a(\gamma|\gamma_0,\tau)$. A lower bound on the variance is now obtained by considering the quantum version of the Van Trees matrix~\cite{Paris2009}, with elements
\begin{equation}
\begin{aligned}    
    V_{i,j} = \int A(\theta,\gamma)Q_{i,j}d\theta d\lambda
    +\frac{1}{N}\int\frac{(\partial_i A(\theta,\gamma)\partial_j A(\theta,\gamma)}{A(\theta,\gamma)} d\theta d\lambda,
\end{aligned}
\end{equation}
instead of the simple QFIM. These are the bounds we shall compare our experimental results to for the SE strategy.

\begin{figure*}[ht!]
    \centering
    \includegraphics[width=0.9\linewidth]{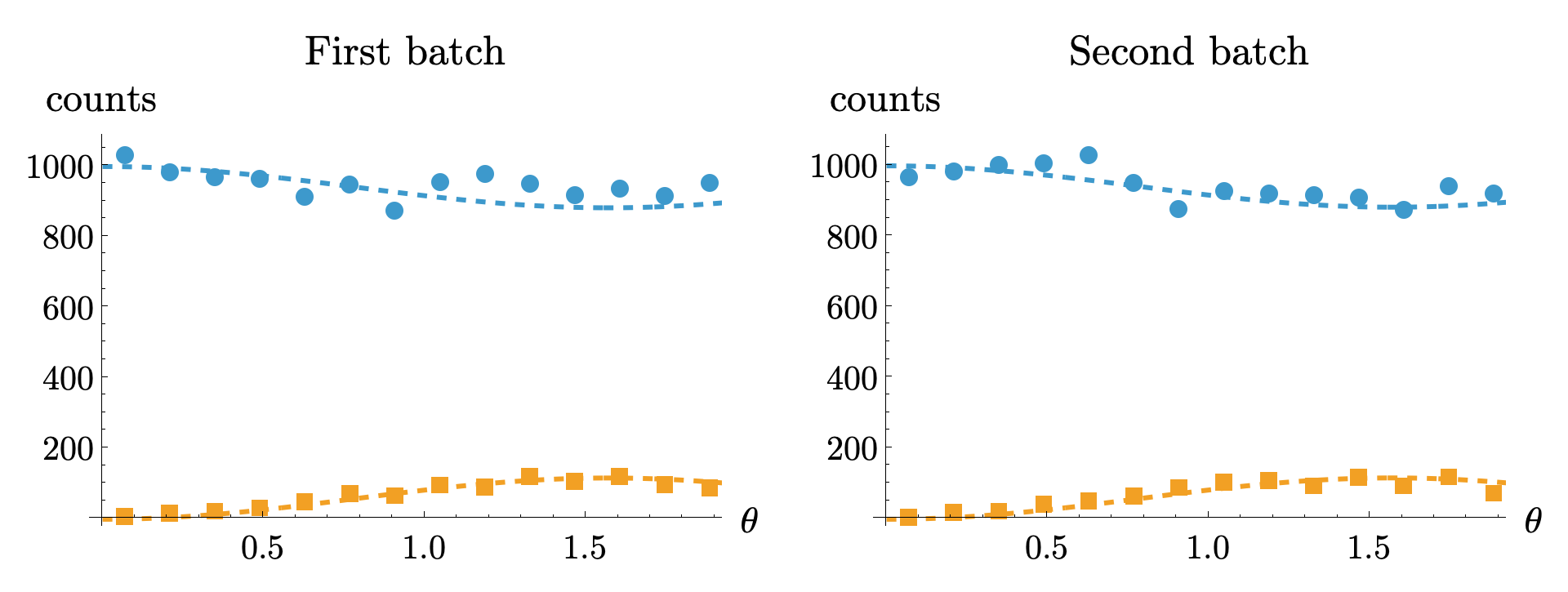}
    \caption{Raw data for the sequential estimation. The counts correspond to measurements along the $Z$ direction: the blue dots correpond to the outcome 0, the golden squares to the outcome 1. The dashed lines are theoretical predictions for $p_0$ and $p_1$ for $\gamma = \pi/9$ as a guide to the eye.}
    \label{fig:experiment}
\end{figure*}

\begin{figure*}[ht!]
    \centering
    \includegraphics[width=\linewidth]{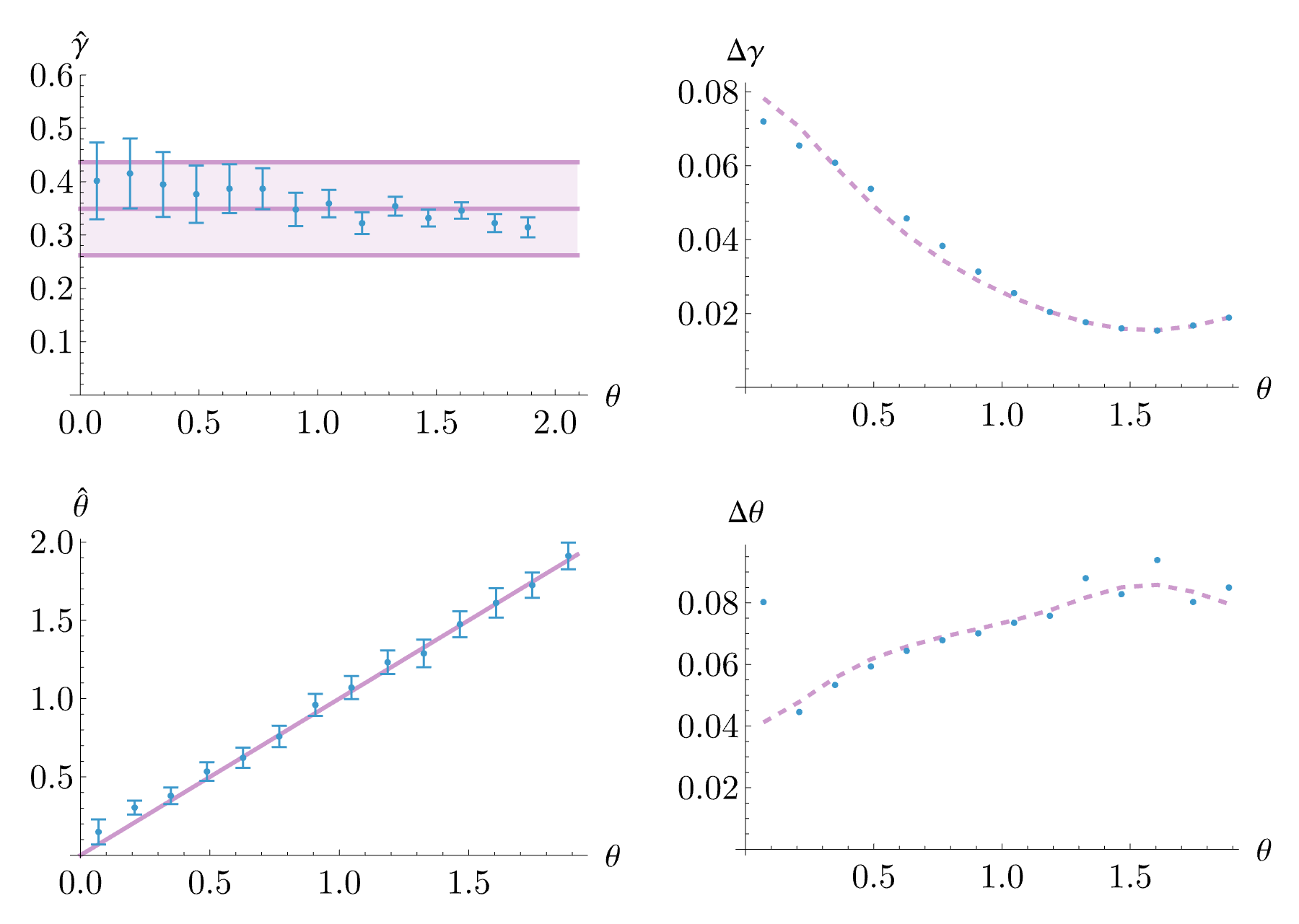}
    \caption{Stepwise two-parameter estimation with prior distributions of width $\tau=5^\circ$. Top left: estimation of the shift $\hat \gamma$ as a function of $\theta$: blue points are for the experimental results, error bars correspond to one standard deviation, while the shaded region indicate the apriori distribution for $\gamma$ (within two stardard deviations). Top right: standard deviation $\Delta\gamma$ as a function of $\theta$. The blue points are the experimental results, the solid line corresponds to the stepwise classical Van Trees limit. Bottom left: results for the estimation of axis $\hat \theta$ as a function of the actual angle $\theta$: the dashed line is the diagonal. Bottom right:  standard deviation $\Delta\theta$ as a function of $\theta$. The blue points are the experimental results, the solid line corresponds to the stepwise Van Trees limit.}
    \label{fig:estimation}
\end{figure*}

The collected counts are reported in Fig.~\ref{fig:experiment} for two repetitions of the experiment. The first one is used for the estimation of $\gamma$, using a Bayesian estimator defined as
\begin{equation}
    \hat \gamma = \int \gamma P_a(\theta|\theta_0,\tau)P_a(\gamma|\gamma_0,\tau)p_0(\gamma,\theta)^{n_0}p_1(\gamma,\theta)^{n_1} d\gamma\,d\theta,
    \label{eq:hatgamma}
\end{equation}
where $n_0$ and $n_1$ are the number of events in either outcome within a detection window. This is the first moment of the distribution for $\gamma$, updated after the measurements. The associated variance is thus the second moment
\begin{equation}
\begin{aligned}
    & \Delta^2 \gamma = \int (\gamma-\hat \gamma)^2 P_a(\theta|\theta_0,\tau)P_a(\gamma|\gamma_0,\tau) \times\\
    &p_0(\gamma,\theta)^{n_0}p_1(\gamma,\theta)^{n_1} d\gamma\,d\theta.
    \end{aligned}
\end{equation}
In the integrals above, we have taken the standard approximation for the binomial as a normal distribution of mean $Np_0(\gamma,\theta)$ and variance $Np_0(\gamma,\theta)p_1(\gamma,\theta)$, as this allows for a more reliable numerical evaluation. Once the estimation of $\gamma$ is obtained, we employ its outcome in defining its distribution in the estimation of the second parameter $\theta$:
\begin{equation}
    \hat \theta = \int \theta P_a(\theta|\theta_0,\tau)P_a(\gamma|\hat \gamma,\Delta\gamma)p_0(\gamma,\theta)^{n'_0}p_1(\gamma,\theta)^{n'_1} d\gamma\,d\theta,  
            \label{eq:hattheta}
\end{equation}
and
\begin{equation}
\begin{aligned}
    & \Delta^2 \theta = \int (\theta-\hat \theta)^2 P_a(\theta|\theta_0,\tau)P_a(\gamma|\hat \gamma,\Delta\gamma) \times\\
    &p_0(\gamma,\theta)^{n_0}p_1(\gamma,\theta)^{n_1} d\gamma\,d\theta,
    \end{aligned}
\end{equation}
where now $n_0'$ and $n_1$ are the number of events in the second batch. We report in Fig.~\ref{fig:estimation} the results for prior distributions with $\tau = 5^\circ$, with $\gamma_0=\pi/9$ for all points, and $\theta_0$ set according to the reading on the rotation mount of the target wave plate. The estimation can be judged satisfactory, and the errors are close to those predicted by the \emph{classical} Van Trees limits for the estimators \eqref{eq:hatgamma} and \eqref{eq:hattheta}, which define the SE strategy. The same procedure can be followed estimating $\theta$ first and then $\gamma$, leading to similar results.

Our measurement strategy based on a single projective measurement would pose difficulties in a JE without adopting a Bayesian framework, as the pertaining classical FIM would be singular. We observe that, in the asymptotic regime, modified bounds can be found by employing the pseudoinverse of the FIM~\cite{Stoica2001,BenHaim2009}.

\begin{figure}[ht!]
    \centering
    \includegraphics[width=\linewidth]{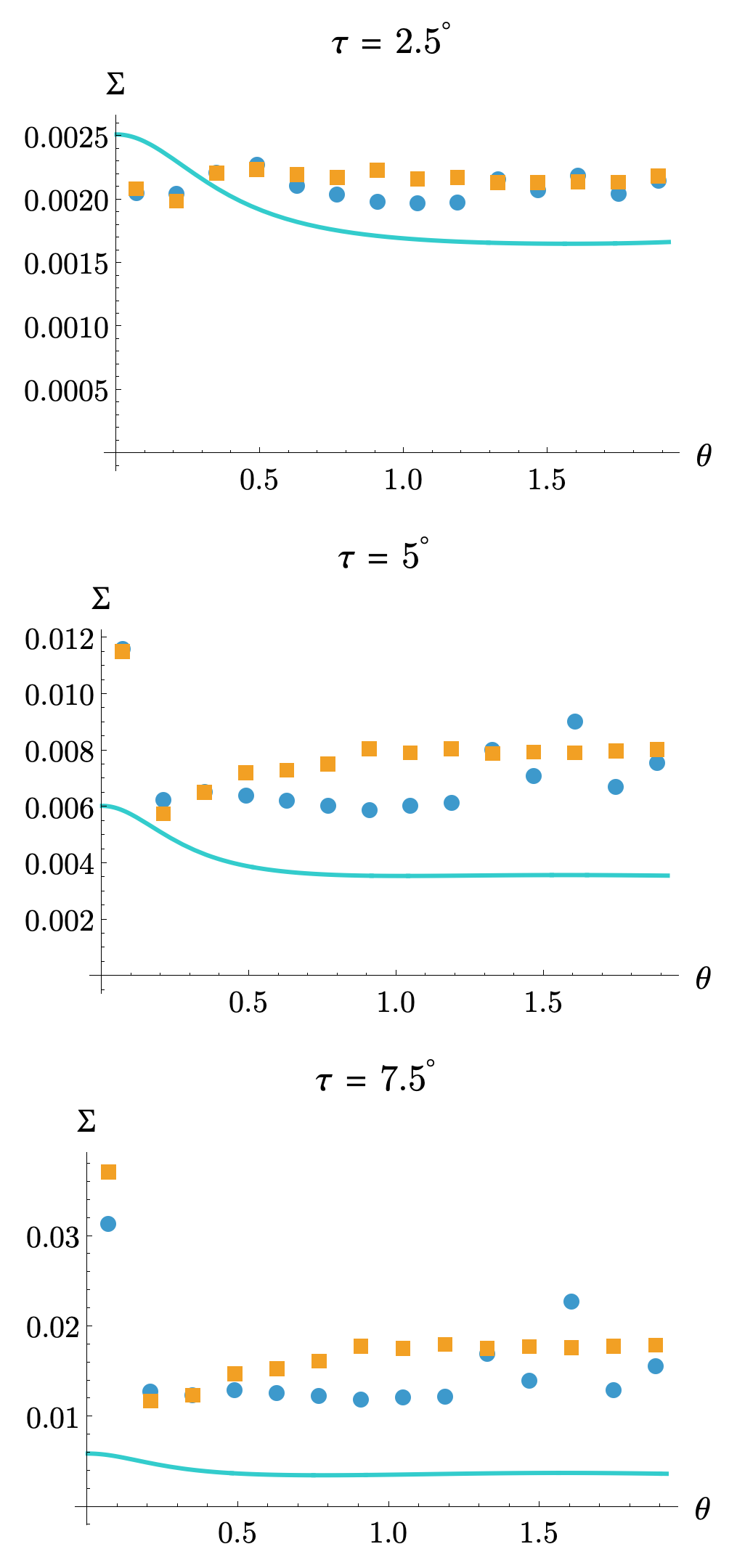}
    \caption{Comparison of the total error $\Sigma$ for different widths of the prior: $\tau =2.5^\circ$ (upper),$\tau =5^\circ$ (middle), 
    $\tau =10^\circ$ (lower). In all panels, the solid line corresponds to the Van Trees limit for JE. Blue points correspond to estimating $\gamma$ first, yellow points to estimating $\theta$ first.}
    \label{fig:comparison}
\end{figure}

A comparison in terms of the total variance $\Sigma=\Delta^2\gamma+\Delta^2\theta$ can be drawn against the quantum Van Trees limit for JE, considering all resources from the two measurement batches. The corresponding results are shown in Fig.~\ref{fig:comparison}, for different widths of the prior distributions, and for both strategies. For $\tau=2.5^\circ$, the observed $\Sigma$ is close to the JE limit, except for the first two points: numerical simulations and comparison with the theoretical stepwise limit reveal these are an artifact due to a slight bias of the estimators. The comparison remains acceptable also for $\tau=5^\circ$, leading us to conclude that the SE in this Bayesian regime can not show an advantage with respect to the joint strategy, but it does not impose too harsh a penalty either. When the a priori is wider, $\tau=10^\circ$, the comparison is less favourable, and it should also be noticed that the experimental variance are further from the classical Van Trees limit; this is a known instance, as this is not a tight bound.    

\section{Discussion and conclusions}

Our results demonstrate that the possible advantage from the stepwise strategy can be spoilt when the Bayesian approach is followed. The reason can be traced in the fact that, while the stepwise Cram\'er-Rao bound is strictly local, in a Bayesian protocol an average is taken over the prior distributions. The final uncertainties are thus affected accordingly. On the other hand, the Bayesian take is most useful in tackling those cases in which the estimators for the individual parameters are mutually dependent, even in the presence of singular Fisher information. It is the case that the \emph{classical} data processing compromises a possible quantum advantage.

Despite these pessimistic observations, the stepwise strategy may still retain a practical advantage since it can work in the presence of simpler measurements: in our example a $Z$ projector is sufficient, whereas the usual joint parameter method would have required a more general POVM. Their implementation and characterisation may be demanding, hence one can trade them for more involved, and careful, data analysis. 

Overall, our results indicate that, although the ultimate quantum advantage of gradual estimation may be reduced in the Bayesian framework, its simplicity and robustness make it a valuable tool for realistic multiparametric quantum metrology, opening the way for its employ in future quantum technologies.

\section*{Acknowledgements}
We thank Valeria Cimini, Marco G. Genoni, Matteo Rosati, and Vittorio Giovannetti for useful discussion.
This work has been realised with the support of  the PRIN project PRIN22-RISQUE-2022T25TR3 of the Italian Ministry of University. G.B. and M.B. are supported by Rome Technopole Innovation Ecosystem (PNRR grant M4-C2-Inv). IG and MB acknowledge the support from MUR Dipartimento di Eccellenza 2023-2027.



\bibliography{biblio.bib}

@article{Paris2009,
  title={Quantum estimation for quantum technology},
  author={Paris, Matteo GA},
  journal={International Journal of Quantum Information},
  volume={7},
  number={supp01},
  pages={125--137},
  year={2009},
  publisher={World Scientific},
  url={https://doi.org/10.1142/S0219749909004839}
}

@article{Liu2020,
doi = {10.1088/1751-8121/ab5d4d},
url = {https://dx.doi.org/10.1088/1751-8121/ab5d4d},
year = {2019},
month = {dec},
publisher = {IOP Publishing},
volume = {53},
number = {2},
pages = {023001},
author = {Liu, Jing and Yuan, Haidong and Lu, Xiao-Ming and Wang, Xiaoguang},
title = {Quantum Fisher information matrix and multiparameter estimation},
journal = {Journal of Physics A: Mathematical and Theoretical}
}

@article{Degen2017,
  title = {Quantum sensing},
  author = {Degen, C. L. and Reinhard, F. and Cappellaro, P.},
  journal = {Rev. Mod. Phys.},
  volume = {89},
  issue = {3},
  pages = {035002},
  numpages = {39},
  year = {2017},
  month = {Jul},
  publisher = {American Physical Society},
  doi = {10.1103/RevModPhys.89.035002},
  url = {https://link.aps.org/doi/10.1103/RevModPhys.89.035002}
}

@article{Giovannetti2006,
	Author = {Giovannetti, Vittorio and Lloyd, Seth and Maccone, Lorenzo},
	Doi = {10.1103/PhysRevLett.96.010401},
	Issn = {0031-9007},
	Journal = {Phys. Rev. Lett.},
	Month = {jan},
	Number = {1},
	Pages = {010401},
	Title = {{Quantum Metrology}},
	Url = {http://link.aps.org/doi/10.1103/PhysRevLett.96.010401},
	Volume = {96},
	Year = {2006}}

@article{Szczykulska2016,
  title={Multi-parameter quantum metrology},
  author={Szczykulska, Magdalena and Baumgratz, Tillmann and Datta, Animesh},
  journal={Advances in Physics: X},
  volume={1},
  number={4},
  pages={621--639},
  year={2016},
  publisher={Taylor \& Francis}
}

@article{Demkowicz2020,
  title={Multi-parameter estimation beyond quantum Fisher information},
  author={Demkowicz-Dobrza{\'n}ski, Rafa{\l} and G{\'o}recki, Wojciech and Gu{\c{t}}{\u{a}}, M{\u{a}}d{\u{a}}lin},
  journal={Journal of Physics A: Mathematical and Theoretical},
  volume={53},
  number={36},
  pages={363001},
  year={2020},
  publisher={IOP Publishing},
  url={https://doi.org/10.1088/1751-8121/ab8ef3}
}

@article{Ragy2016,
  title={Compatibility in multiparameter quantum metrology},
  author={Ragy, Sammy and Jarzyna, Marcin and Demkowicz-Dobrza{\'n}ski, Rafa{\l}},
  journal={Physical Review A},
  volume={94},
  number={5},
  pages={052108},
  year={2016},
  publisher={APS},
  url={https://doi.org/10.1103/PhysRevA.94.052108?_gl=1*119wno7*_ga*MTg3MDc5ODcyLjE3NDI5MDEyNTk.*_ga_ZS5V2B2DR1*MTc0MzAxMjc3OS42LjEuMTc0MzAxMzYwNC4wLjAuMTExOTc0MDQ3MA..}
}

@article{Albarelli2020,
  title={A perspective on multiparameter quantum metrology: From theoretical tools to applications in quantum imaging},
  author={Albarelli, Francesco and Barbieri, Marco and Genoni, Marco G and Gianani, Ilaria},
  journal={Physics Letters A},
  volume={384},
  number={12},
  pages={126311},
  year={2020},
  publisher={Elsevier},
  url={https://doi.org/10.1016/j.physleta.2020.126311}
}

@article{Waterfall2006,
  title = {Sloppy-Model Universality Class and the Vandermonde Matrix},
  author = {Waterfall, Joshua J. and Casey, Fergal P. and Gutenkunst, Ryan N. and Brown, Kevin S. and Myers, Christopher R. and Brouwer, Piet W. and Elser, Veit and Sethna, James P.},
  journal = {Phys. Rev. Lett.},
  volume = {97},
  issue = {15},
  pages = {150601},
  numpages = {4},
  year = {2006},
  month = {Oct},
  publisher = {American Physical Society},
  doi = {10.1103/PhysRevLett.97.150601},
  url = {https://link.aps.org/doi/10.1103/PhysRevLett.97.150601}
}

@article{Goldberg2021,
  title={Taming singularities of the quantum Fisher information},
  author={Goldberg, Aaron Z and Romero, Jos{\'e} L and Sanz, {\'A}ngel S and S{\'a}nchez-Soto, Luis L},
  journal={International Journal of Quantum Information},
  volume={19},
  number={08},
  pages={2140004},
  year={2021},
  publisher={World Scientific}
}

@article{Mukhopadhyay2025,
  title={Beating joint quantum estimation limits with stepwise multiparameter metrology},
  author={Mukhopadhyay, Chiranjib and Bayat, Abolfazl and Montenegro, Victor and Paris, Matteo GA},
  journal={arXiv preprint arXiv:2506.06075},
  year={2025}
}

@article{He2025,
  title={Scrambling for precision: optimizing multiparameter qubit estimation in the face of sloppiness and incompatibility},
  author={He, Jiayu and Paris, Matteo GA},
  journal={arXiv preprint arXiv:2503.08235},
  year={2025}
}

@article{Rozema2014,
  title = {Optimizing the Choice of Spin-Squeezed States for Detecting and Characterizing Quantum Processes},
  author = {Rozema, Lee A. and Mahler, Dylan H. and Blume-Kohout, Robin and Steinberg, Aephraim M.},
  journal = {Phys. Rev. X},
  volume = {4},
  issue = {4},
  pages = {041025},
  numpages = {9},
  year = {2014},
  month = {Nov},
  publisher = {American Physical Society},
  doi = {10.1103/PhysRevX.4.041025},
  url = {https://link.aps.org/doi/10.1103/PhysRevX.4.041025}
}

@article{Zhou2015,
author = {Xiao-Qi Zhou and Hugo Cable and Rebecca Whittaker and Peter Shadbolt and Jeremy L. O'Brien and Jonathan C. F. Matthews},
journal = {Optica},
number = {6},
pages = {510--516},
publisher = {Optica Publishing Group},
title = {Quantum-enhanced tomography of unitary processes},
volume = {2},
month = {Jun},
year = {2015},
url = {https://opg.optica.org/optica/abstract.cfm?URI=optica-2-6-510},
doi = {10.1364/OPTICA.2.000510},
}

@ARTICLE{Stoica2001,
  author={Stoica, P. and Marzetta, T.L.},
  journal={IEEE Transactions on Signal Processing}, 
  title={Parameter estimation problems with singular information matrices}, 
  year={2001},
  volume={49},
  number={1},
  pages={87-90},
  doi={10.1109/78.890346}}

@ARTICLE{BenHaim2009,
  author={Ben-Haim, Zvika and Eldar, Yonina C.},
  journal={IEEE Signal Processing Letters}, 
  title={On the Constrained CramÉr–Rao Bound With a Singular Fisher Information Matrix}, 
  year={2009},
  volume={16},
  number={6},
  pages={453-456},
  doi={10.1109/LSP.2009.2016831}}

@misc{fazio2025orders,
      title={Orders matter: tight bounds on the precision of sequential quantum estimation for multiparameter models}, 
      author={Gabriele Fazio and Jiayu He and Matteo G. A. Paris},
      year={2025},
      eprint={2510.14963},
      archivePrefix={arXiv},
      primaryClass={quant-ph},
      url={https://arxiv.org/abs/2510.14963}, 
}

\end{document}